\documentclass{ws-procs9x6}

\begin{document}

\count255=\time\divide\count255 by 60 \xdef\hourmin{\number\count255}
  \multiply\count255 by-60\advance\count255 by\time
 \xdef\hourmin{\hourmin:\ifnum\count255<10 0\fi\the\count255}

\newcommand{\xbf}[1]{\mbox{\boldmath $ #1 $}}

\newcommand{\sixj}[6]{\mbox{$\left\{ \begin{array}{ccc} {#1} & {#2} &
{#3} \\ {#4} & {#5} & {#6} \end{array} \right\}$}}

\newcommand{\threej}[6]{\mbox{$\left( \begin{array}{ccc} {#1} & {#2} &
{#3} \\ {#4} & {#5} & {#6} \end{array} \right)$}}

\newcommand{\clebsch}[6]{\mbox{$\left( \begin{array}{cc|c} {#1} & {#2} &
{#3} \\ {#4} & {#5} & {#6} \end{array} \right)$}}

\newcommand{\iso}[6]{\mbox{$\left( \begin{array}{cc||c} {#1} & {#2} &
{#3} \\ {#4} & {#5} & {#6} \end{array} \right)$}}

\title{$\pi N \to$ Multi-$\pi N$ Scattering in the $1/N_c$ Expansion}

\author{Herry J. Kwee$^*$}

\address{Physics Department, Arizona State University,\\
Tempe, AZ 85287-1504, USA\\
$^*$E-mail: Herry.Kwee@asu.edu}

\begin{abstract}
We extend the $1/N_c$ meson-baryon scattering formalism to $\pi N \to$
multi-$\pi N$ case.  We first show that the leading-order large $N_c$
processes proceed through resonant intermediate states (e.g., $\rho N$ or
$\pi \Delta$).  We find that the pole structure of baryon resonances can
be uniquely identified by their (non)appearance in $\eta N$ or mixed
partial-wave $\pi \Delta$ final states.
\end{abstract}


\bodymatter

\section{Introduction} \label{intro}
This talk mainly focuses on $\pi N \! \to$ multi-$\pi N$ processes in the
$1/N_c$ scattering formalism~\cite{paper}.  Recent development of the
original scattering
formalism~\cite{CLcompat,CLdecouple,CL1,CL2,CDLN,CDLM,ItJt} can be
reviewed in Refs.~\cite{confs}.  The central idea rests
upon symmetries that emerge for QCD in the large $N_c$ limit.  The
symmetries relate scattering amplitudes in channels of different $I$, $J$,
and other quantum numbers and consequently impose degeneracies among poles
that occur within them.

\begin{figure}[ht]
\epsfxsize 1.3 in \epsfbox{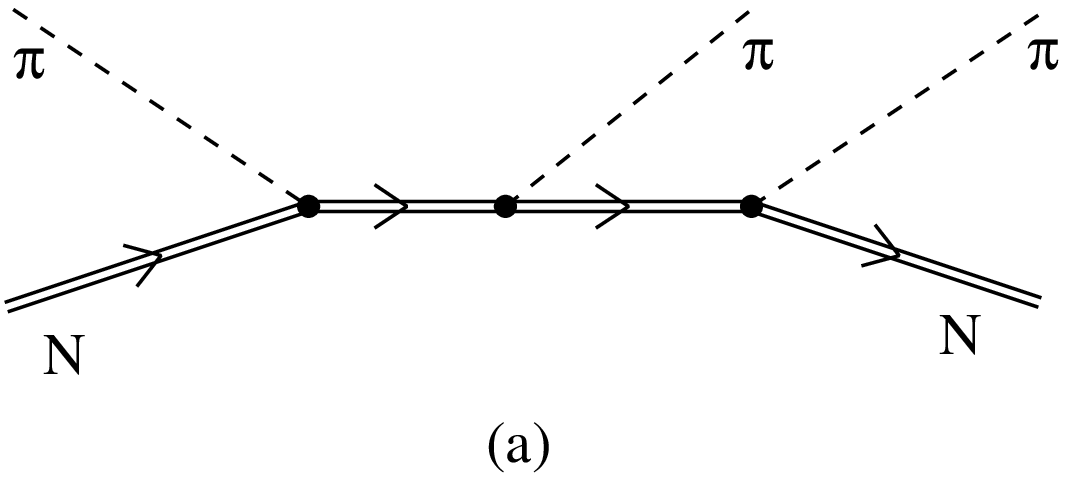} \hspace{1em}
\epsfxsize 1.3 in \epsfbox{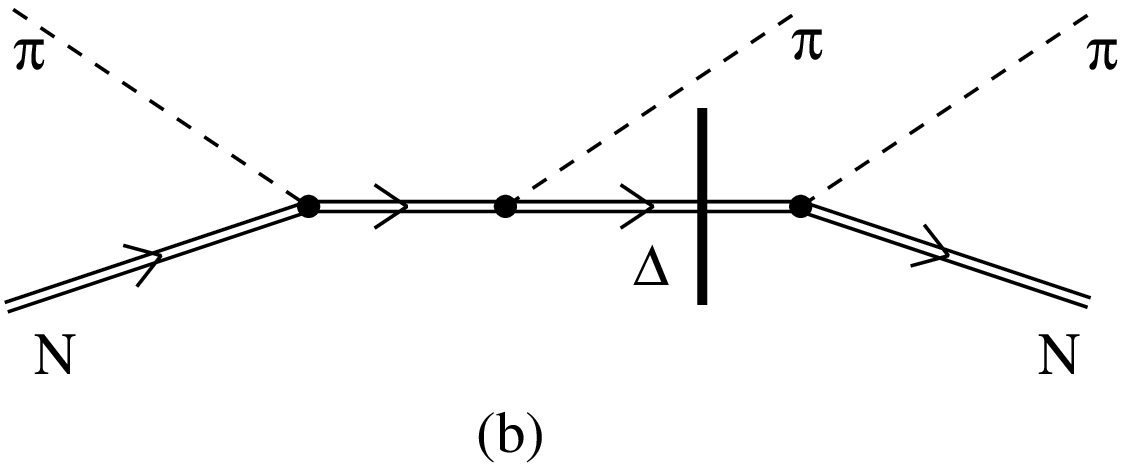} \hspace{1em}
\epsfxsize 1.3 in \epsfbox{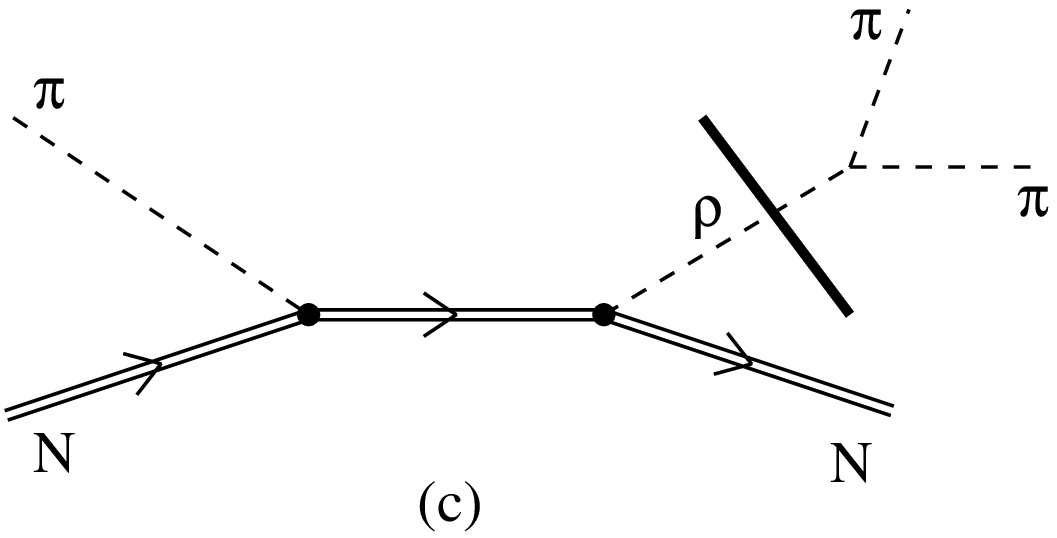}
\caption{Diagrams for $\pi N \! \to \! \pi \pi N$ scattering.  (a)
Nonresonant scattering (1 of 6 diagrams); (b) $\pi N \! \to \! \pi
\Delta$ ($\Delta$ on shell), followed by $\Delta \! \to \! \pi N$; (c)
$\pi N \! \to \! \rho N$, ($\rho$ on shell) followed by $\rho \! \to
\! \pi \pi$.}
\label{scatfigs}
\end{figure}

However, this scattering formalism depends upon a single incoming source
and a single outgoing source scattering from the baryon.  Therefore, a
meaningful way to constrain such multipion processes must be considered.
Also, a standard $N_c$ counting shows that the generic scattering
amplitude for $\pi N \! \to \! \pi N$ is $O(N_c^0)$, while that for $\pi N
\! \to \! \pi \pi N$ [Fig.~\ref{scatfigs}(a)] is $O(N_c^{-1/2})$.
Nevertheless, circumstances exist in which processes that {\em
eventually\/} produce two (or more) pions nevertheless appear with
amplitudes at leading order, $O(N_c^0)$.  In particular, the $\Delta$ is
stable for sufficiently large $N_c$; its width scales as $1/N_c^2$. 
Therefore, as in Fig.~\ref{scatfigs}(b), the $\pi N \! \to \! \pi \pi N$
process may be cut (indicating an on-shell state) at the intermediate
stage, $\pi N \! \to \! \pi \Delta$.  Of course, we live in the $N_c \! =
\! 3$ world where $\Gamma_\Delta$ is over 100~MeV; even so,
$\Gamma_\Delta$ is considered sufficiently small that researchers
regularly extract $\pi N \! \to \! \pi \Delta$ partial widths.  Similarly,
processes such as in Fig.~\ref{scatfigs}(c) ($\pi N \! \to \! \rho N$
followed by $\rho \! \to \! \pi \pi$) may be analyzed in the two-body
formalism since $\pi N \! \to \! \rho N$ partial widths have been
tabulated.

\section{Group Theory Preliminaries} \label{group}

The derivation of the meson-baryon scattering amplitudes
formalism~\cite{HEHW,MM,Mattis} is done by considering only baryons
lying in the ground-state band of large $N_c$, whose nonstrange members
have spin = isospin $R \! = \! \frac 1 2, \frac 3 2,\ldots,
\frac{N_c}{2}$.  This restriction is not physically constraining since all
observed meson-baryon scattering processes fit into this category.  The
basic process is
\begin{equation}
m + B \to m^\prime + B^\prime ,
\end{equation}
where $m \, (m^\prime)$ is a meson of spin $s \, (s^\prime)$ and
isospin $i \, (i^\prime)$, in a state of relative orbital angular
momentum $L$ ($L^\prime$) with a baryon $B \, (B^\prime)$ of spin =
isospin $R \, (R^\prime)$ in the ground-state multiplet, and the
total spin angular momentum of the meson and baryon is denoted $S \,
(S^\prime)$.  The intermediate state is labeled by quantum numbers $I$ and
$J$, giving the full partial wave $S_{LL^\prime SS^\prime IJ}$. 
Abbreviating the multiplicity $2X\!+\!1$ of an SU(2) representation of
quantum number $X$ by $[X]$, one finds~\cite{MM,CLcompat}
\begin{eqnarray}
S_{L L^\prime S S^\prime I J} & = & \sum_{K, \tilde{K} ,
\tilde{K}^\prime} [K]
([R][R^\prime][S][S^\prime][\tilde{K}][\tilde{K}^\prime])^{1/2}
\nonumber \\
& & \times \left\{ \begin{array}{ccc}
L & i & \tilde{K} \\
S & R & s \\
J & I & K \end{array} \right\}
\left\{ \begin{array}{ccc}
L^\prime & i^\prime & \tilde{K}^\prime \\
S^\prime & R^\prime & s^\prime \\
J & I & K \end{array} \right\}
\tau_{K \tilde{K} \tilde{K}^\prime L L^\prime} . \label{Mmaster}
\end{eqnarray}
The remaining symbols, $K$, $\tilde{K}$, and $\tilde{K}^\prime$, are
intermediate quantum numbers, with ${\bf K} \! \equiv \! {\bf I} \! + \!
{\bf J}$, $\tilde {\bf K} \! \equiv \! {\bf i} \! + \! {\bf L}$, and
$\tilde {\bf K}^\prime \! \equiv \!  {\bf i^\prime} \!  + \!  {\bf
L}^\prime$ (so that ${\bf K} \! = \! \tilde {\bf K} \! + \!  {\bf s}
\! = \! \tilde {\bf K}^\prime \!  + \! {\bf s^\prime}$).  The factors in
braces are $9j$ coefficients, and $\tau_{K\tilde{K}\tilde{K}^\prime
LL^\prime}$ are universal amplitudes ({\it reduced\/} or {\it K\/}
amplitudes) that are independent of $I$, $J$, $R$, $R'$, $i$, $i^\prime$,
$s$, and $s^\prime$.  The linear relations among the scattering amplitudes
can be seen from the structure of Eq.~(\ref{Mmaster}); the point is there
are more $S_{LL^\prime S S^\prime IJ}$ amplitudes than
$\tau_{K\tilde{K}\tilde{K}^\prime LL^\prime}$ amplitudes.

\section{Amplitude Tables} \label{result}

Here we present the transition amplitude for $\eta N$, $\eta \Delta$, $\pi
N$, $\pi \Delta$, $\omega N$, and $\rho N$ final states of spin $\frac 1
2$, $I \! = \! \frac 1 2$ and both positive-${\cal P}$ and negative-${\cal
P}$ parity resonances in Tables~\ref{t1} and~\ref{t2}.  The notation for
$\pi N \! \to \! m^\prime B^\prime$ partial waves is $LL^{\prime \; (\pi
N) (m^\prime B^\prime)_{S^\prime}}_{2I, 2J}$; if $L^\prime \! = \! L$ then
the label $L^\prime$ is suppressed, while if $s^\prime$ (the $m^\prime$
spin) is zero, then $S^\prime$ equals the spin of baryon $B^\prime$ and is
suppressed.

\begin{table}
\tbl{Partial-wave amplitudes for positive-parity $N_{1/2}$ resonances in
multipion processes (the $\pi N$ final state is included for
comparison).  Expansions are given in terms of $K$ amplitudes.
\label{t1}}
{\begin{tabular}{lcccccl}
State \mbox{  } && Poles \mbox{   } &&
\multicolumn{3}{l}{Partial Wave, $K$-Amplitudes} \\
\hline\hline
$N_{1/2}^{+}$ && $K = 0, 1$ &&
$P^{{(\pi N)} {(\eta N)}}_{11}$
&=& $-\frac{\sqrt 2}{\sqrt3}\tau_{11111}$ \\
&& && $P^{{(\pi N)} {(\pi N)}}_{11}$
&=& $\frac{1}{3}\tau_{00011} + \frac{2}{3}\tau_{11111}$ \\
&& && $P^{{(\pi N)} {(\pi \Delta)}}_{11}$
&=& $\frac{\sqrt 2}{3}\tau_{00011} - \frac{\sqrt 2}{3} \tau_{11111}$ \\
&& && $P^{{(\pi N)} {(\omega N)}_1}_{11}$
&=& $\frac{1}{3}\tau_{00111} +\frac{2}{3} \tau_{11111}$ \\
&& && $P^{{(\pi N)} {(\omega N)}_3}_{11}$
&=& $\frac{\sqrt 2}{3}\tau_{00111} -\frac{\sqrt 2}{3} \tau_{11111}$ \\
&& && $P^{{(\pi N)} {(\rho N)}_1}_{11}$
&=& $\frac{\sqrt 2}{3\sqrt 3}\tau_{00111} - \frac{\sqrt 2}{9} \tau_{11011} +
\frac{2\sqrt{10}}{9}\tau_{11211}$\\
&& && $P^{{(\pi N)} {(\rho N)}_3}_{11}$
&=& $-\frac{1}{3\sqrt 3}\tau_{00111} - \frac{4}{9}\tau_{11011}$ \\
&& && &&$+\frac{1}{\sqrt 3}\tau_{11111} + \frac{\sqrt 5}{9}\tau_{11211}$ \\
\hline
\end{tabular}}
\end{table}

\begin{table}
\tbl{Partial-wave amplitudes for negative-parity $N_{1/2}$ resonances in
multipion processes.
\label{t2}}
{\begin{tabular}{lcccccl}
State \mbox{  } && Poles \mbox{   } &&
\multicolumn{3}{l}{Partial Wave, $K$-Amplitudes} \\
\hline\hline
$N_{1/2}^{-}$ && $K = 1$ &&
$S^{{(\pi N)} {(\eta N)}}_{11}$
&=& $0$ \\
&& && $S^{{(\pi N)} {(\pi N)}}_{11}$
&=& $\tau_{11100}$ \\
&& && $SD^{{(\pi N)} {(\pi \Delta)}}_{11}$
&=& $- \tau_{11102}$ \\
&& && $S^{{(\pi N)} {(\omega N)}_1}_{11}$
&=& $\tau_{11000}$ \\
&& && $SD^{{(\pi N)} {(\omega N)}_3}_{11}$
&=& $- \tau_{11202}$ \\
&& && $S^{{(\pi N)} {(\rho N)}_1}_{11}$
&=& $\sqrt{\frac{2}{3}}\tau_{11100}$\\
&& && $SD^{{(\pi N)} {(\rho N)}_3}_{11}$
&=& $\frac{1}{\sqrt 6}\tau_{11102} + \frac{1}{\sqrt 2}\tau_{11202}$ \\
\hline
\end{tabular}}
\end{table}

\section{Phenomenological Results} \label{phenom}

The association of resonances with poles---as determined by presence of
absence of certain decay channels---seems robust.  In particular,
Eq.~(\ref{Mmaster}) can be employed in a straightforward fashion to show
that $\pi N \! \to \! \eta N$ contains a single $K$ amplitude [with $K \!
= \! L$], and the mixed partial wave $\pi N (L) \! \to \pi \Delta
(L^\prime)$ contains a single $K$ amplitude [with $K \! = \! \frac 1 2 (L
\! + \! L^\prime)$]~\cite{CLdecouple}.  For given $I$, $J$, and ${\cal P}$
these two amplitudes always probe distinct $K$.  The following is an
example of brief analysis for one of the two channels.

\begin{enumerate}

\item $\bm{N^{+}_{1/2} \: (P_{11})}$:
The two well-established resonances in this channel are $N(1440)$ (the
Roper) and $N(1710)$, while our transition amplitude calculations provide
two distinct pole structures, $K \! = \! 0$ and $K \! = \! 1$.  The
$N(1440)$ has a very small, $(0 \! \pm \!  1) \%$, $\eta N$ BR while
$N(1710)$ has a small but nonnegligible $\eta N$ BR, $(6.2\pm 1.0)\%$. 
Comparing this observation to our tabulated result for the $\pi N \! \to
\! \eta N $ transition amplitude suggests that the Roper is a $K \! = \! 
0$ pole and the $N(1710)$ is a $K \! = \! 1$ pole.  This assignment agrees
very well with the assumption of the Roper as a radial excitation of
ground-state $N$, which is a (nonresonant) $K \!  = \!  0$ state.

\end{enumerate}

\section*{Acknowledgments}
The authors thank Tom Cohen for valuable discussions.  This work was
supported by the NSF under Grant No.\ PHY-0456520.

\end{document}